\documentclass[preprint,showpacs,preprintnumbers,nofootinbib]{revtex4}

\usepackage{graphicx}
\usepackage{dcolumn}
\usepackage{bm}

\newcommand{\bea}{\begin{eqnarray}}
\newcommand{\eea}{\end{eqnarray}}
\newcommand{\be}{\begin{equation}}
\newcommand{\ee}{\end{equation}}
\newcommand{\pp}{\psi^{\prime}}
\newcommand{\xp}{\xi^{\prime}}
\newcommand{\cp}{\chi^{\prime}}
\newcommand{\fp}{\phi^{\prime}}
\newcommand{\vp}{\varphi^{\prime}}
\newcommand{\lpp}{\lambda^{\prime \prime}}
\newcommand{\npp}{\nu^{\prime \prime}}
\newcommand{\np}{\nu^{\prime}}
\newcommand{\lp}{\lambda^{\prime}}
\newcommand{\lps}{\lambda^{\prime 2}}
\newcommand{\nps}{\nu^{\prime 2}}
\newcommand{\gn}{8 \pi M_{p}^{-2}}
\newcommand{\ppp}{\psi^{\prime\prime}}
\newcommand{\xpp}{\xi^{\prime\prime}}
\newcommand{\cpp}{\chi^{\prime\prime}}

\newcommand{\vpp}{\varphi^{\prime\prime}}

\newcommand{\ch}{{\cal{H}}}
\begin{document}

\preprint{BROWN-HET-1382}

\title{Linear Perturbations in Brane Gas Cosmology}

\author{Scott Watson}
 \email{watson@het.brown.edu}
\author{Robert Brandenberger}%
 \email{rhb@het.brown.edu}
\affiliation{Department of Physics, Brown University, Providence, RI.}

\date{\today}

\begin{abstract}
We consider the effect of string inhomogeneities on the time
dependent background of Brane Gas Cosmology.  We derive the equations
governing the linear perturbations of the dilaton-gravity background
in the presence of string matter sources. We focus on long wavelength
fluctuations and find that there are no instabilities. Thus, the
predictions of Brane Gas Cosmology are robust against the introduction
of linear perturbations. In particular, we find that the stabilization of
the extra dimensions (moduli) remains valid in the presence of dilaton and
string perturbations.
\end{abstract}

\pacs{Valid PACS appear here}
\maketitle

\section{Introduction}

Understanding the behavior of strings in a time dependent background
has been a subject of much interest and has been pursued in a number of
differing ways. One scenario, known as Brane Gas Cosmology (BGC) is devoted to
understanding the effect that string and brane gases could have on a
dilaton-gravity background in the early Universe
\cite{bv,vafa,bgc,isotropization,stable,extended}.
In \cite{bv}, it was suggested that the energy
associated with the winding of strings around the compact dimensions
would produce a confining potential for the
scale factor and halt the cosmological expansion\footnote{This was
later shown quantitatively in \cite{vafa}.}. The analysis of BGC was
initially performed at the level of the equations for homogeneous
and isotropic cosmology. The results were recently
extended to the case of anisotropic cosmology in \cite{isotropization}.
There, it was shown that string gases can result in three dimensions
growing large and isotropic due
to string annihilation while the other six dimensions remain confined.
In \cite{stable} it was shown that by considering both momentum and
winding modes of strings,
the six confined dimensions can be stabilized at the self-dual radius,
where the energy of the string gas is minimal.
This result demonstrated that, in BGC, the volume moduli of the extra
dimensions can be stabilized in a natural and intuitive way.

To date, however, all analyses have been performed at the level of the
equations for homogeneous cosmology. The string sources
are usually represented by a perfect fluid with homogeneous energy and
pressure densities given by the mass spectrum of the strings
(see e.g. \cite{bgc,stable}).
One may worry that inhomogeneities
of string sources (in particular strings winding around the confined
dimensions) as a function of the unconfined spatial directions could
lead to serious instabilities which would ruin one of the main
successes of BGC, namely the prediction that three directions become
large, leaving the other six confined uniformly as a function of the
coordinates of the large spatial sections.

As a first step towards addressing this concern, we here study the
evolution of linear fluctuations about a BGC background. We introduce
linear matter and metric fluctuations, derive the linear perturbation
equations, and study their stability on long wavelength (wavelengths
larger than the Hubble radius) \footnote{On shorter wavelengths, we
expect that the motion of the strings will smear out potential
instabilities in a way analogous to how the motion of light particles
(``free-streaming'') leads to a decay of short wavelength fluctuations
in standard cosmology (see e.g. \cite{Peebles} for a review).} Note
that since the dilaton plays a vital role in the stabilization process,
we must also consider the effect of inhomogeneities in the dilaton field.
Our main result is that the long wavelength fluctuations do not grow
in time - there are no instabilities at the level of a linearized
analysis.

In Section 2, we present the equations of BGC and review the background
solution that leads to $3+1$ dimensions growing large and $6$ dimensions
being stabilized at the self-dual radius.
In Section 3, we introduce the fluctuations and derive the
equations for linearized perturbations about the time dependent homogeneous
background of BGC. The long wavelength solutions
are given in Section 4, leaving a detailed description for the Appendix.
We conclude with a discussion of the results in Section 5.

\section{Background Solution}

Our starting point is the low energy effective action for the bulk space-time
with string matter sources \cite{vafa},
\be
\label{action}
S=\frac{1}{4\pi \alpha^{\prime}}\int d^{D}x \sqrt{-g}
e^{-2 \varphi}\Bigl( R+4(\nabla \varphi)^{2}-\frac{1}{12}H^{2}
\Bigr)+S_{\text m} \, ,
\ee
where $R$ denotes the Ricci scalar, $g$ is the determinant of the
background metric, $\varphi$ is the dilaton field, and $H$ is the field
strength of an antisymmetric tensor field. The action of the matter
sources is denoted by $S_{\text m}$.
For example, with $D=10$ this is the low energy effective action of
type II-A superstring theory.
For the purposes of this paper we will ignore the effects of
branes, since it will be the winding and momentum modes of the string that
ultimately determine the dimensionality and stability of
space-time \cite{bgc}. Here, we will ignore the effects of
fluxes \footnote{See \cite{Campos:2003ip} for inclusion of
fluxes in the scenario.}, i.e. we set $H=0$.

This action yields the following equations of motion,
\bea
\label{eomb}
& & R_{\mu}^{\; \; \nu} +2 \nabla_{\mu} \nabla^{\nu}\varphi
=8 \pi M^{-2}_{p} \; e^{2\varphi} T_{\mu}^{\; \; \nu}, \nonumber \\
& & R+4\nabla_{\kappa}\nabla^{\kappa}\varphi
-4\nabla_{\kappa}\varphi \nabla^{\kappa} \varphi=0,
\eea
where $\nabla$ is the covariant derivative.

We will work in the conformal frame with a homogeneous metric of the form
\be \label{metric}
ds^{2}=e^{2 \lambda(\eta)} \Bigl( d\eta^{2}- \delta_{i j} dx^{i} dx^{j}\Bigr)
- e^{2 \nu(\eta)} \delta_{m n} dx^{m}dx^{n},
\ee
where $(\eta,x^i)$ are the coordinates of $3+1$ space-time
and $x^m$ are the coordinates of the other six dimensions, both of which
can be
taken to be isotropic \cite{isotropization}.
The scale factors $a(\eta)$ and $b(\eta)$ are given by $\lambda \equiv
\ln(a)$ and $\nu \equiv \ln(b)$.

We consider the effect of the strings on the background through their
stress energy tensor
\be \label{stress}
T_{\mu}^{\: \: \nu} \equiv diag(\rho,-p_{i},-p_{m}),
\ee
where $\rho$ is the energy density of the strings, $p_i$ ($i=1 \ldots 3$) is
the pressure in the expanding dimensions and
$p_m$ ($m=4 \ldots 9$) is the pressure in the small dimensions (because
of our assuption of isotropy of each subspace, there is only one independent
$p_i$ and one independent $p_m$).

Strings contain winding modes, momentum modes and oscillatory modes.
However, since the energies of the oscillatory modes are independent
of the size of the dimensions, and since the winding modes and
momentum modes dominate the thermodynamic partition function at very
small and very large radii of the spatial dimensions, we shall here
neglect the oscillatory modes. In the absence of string interactions,
the contributions to the stress tensor coming from the string
winding modes and momentum
modes ($T^{\text{w}}_{\mu\nu}$ and $T^{\text{m}}_{\mu\nu}$ respectively)
are separately conserved,
\bea
T_{\mu\nu}=T^{\text{w}}_{\mu\nu}+T^{\text{m}}_{\mu\nu} \nonumber\\
\nabla^{\mu} T^{\text{w}}_{\mu\nu}=0 \;\;\;\;\;\;
\nabla^{\mu} T^{\text{m}}_{\mu\nu} \label{conservationeq}=0 \, .
\eea
The conservation equations take the form
\be \label{theconservationeq}
\rho^{\prime \: {\text w,m}} +\sum_{i=1}^9 \lambda_i^{\prime}
(\rho^{\text w,m} - p^{\text w,m}_i)=0,
\ee
where the derivatives are with respect to the conformal time $\eta$, and
where for the moment we consider 9 independent scale factors.

Expressing (\ref{eomb}) in terms of the metric (\ref{metric}) and the
stress tensor (\ref{stress}),
we find the following system of equations,
\bea \label{theset}
-3 \lpp -6 \npp +6 \lp \np -6\nps +\vpp-\lp \vp=\gn
e^{\varphi+2\lambda}\rho\\
-\lpp+2\lps+6\lp\np+\lp\vp=-\gn e^{\varphi+2\lambda}p_{i}\\
-\npp+6\nps+2\lp\np+\vp\np=-\gn e^{\varphi+2\lambda}p_{m} \label{it}\\
-6\lpp-12\npp-24\lp\np-42\nps -6\lps -\varphi^{\prime
2}+2\vpp+8\lp\vp+12\vp\np=0. \label{set}
\eea

The explicit forms of the energy density and pressure were given in
\cite{stable}\footnote{The equations here are related to Eq. (18)
in \cite{stable} by the volume factor $V=e^{3\lambda+6\nu}$, e.g.
$\rho=\frac{E}{V}$},
\bea \label{sources1}
\rho=3 \mu N^{(3)} e^{-2 \lambda-6 \nu}+3 \mu M^{(3)} e^{-4 \lambda-6 \nu}+
6 \mu N^{(6)} e^{-3 \lambda-5 \nu}+6 \mu M^{(6)} e^{-3 \lambda-7 \nu},\\
p_i=-\mu N^{(3)} e^{-2 \lambda-6 \nu}+\mu M^{(3)} e^{-4 \lambda-6 \nu}, \\
p_m=-\mu N^{(6)} e^{-3 \lambda-5 \nu}+\mu M^{(6)} e^{-3 \lambda-7 \nu}
\label{sources1b},
\eea
where $\mu$ is a constant, $N^{(3)}$ and $M^{(3)}$ are the numbers of
winding and momentum modes in the large directions, and $N^{(6)}$ and
$M^{(6)}$ in the six small directions.

We are interested in solutions that stabilize the internal
dimensions, while allowing the three large dimensions to expand.
Such solutions were discussed in \cite{stable}, where it was shown
that the winding and momentum modes of the strings lead naturally to
stable compactifications of the internal dimensions at the self dual radius,
while the other three dimensions grow large since the string modes
winding them have annihilated, allowing for expansion. Thus, we
will set $N^{(3)}=0$.
At the self dual radius, the number of winding modes is equal to the number
of momentum modes (i.e. $N^{(6)}=M^{(6)}$) and the pressure vanishes
($p_{m}=0$).

In Ref. \cite{stable}, the solutions subject to the above
conditions on the winding and momentum numbers were found numerically.
In this paper, we wish to study the stability of these solutions towards
linear perturbations in the time interval when
the internal dimensions have stabilized and
the large dimensions give power law expansion. In the following section,
we will derive the equations for the linear fluctuations. The coefficients
in these equations depend on the background solution. We will
use analytical expressions which approximate the numerically obtained
solutions of \cite{stable}. We restrict our
initial conditions so that the evolution preserves the low energy
and small string coupling assumptions ($g_s \sim e^{2 \varphi } \ll 1$).

A approximation to a typical solution of
the equations (\ref{theset}-\ref{set}) is found to be of the form
\bea
\lambda(\eta)= k_1 \ln(\eta) + \lambda_{0} \;\;\; {\text or} \;\;\; a(\eta) = a_{0} \eta^{k_1}, \nonumber \\
\varphi(\eta)=-k_2 \ln(\eta) + \varphi_{0},
\label{backsoln}
\eea
where the constants $k_1$, $k_2$, $\lambda_{0}$ and $\varphi_{0}$
depend on the choice of initial conditions.
We have made use of $\nu=\nu^{\prime}=\nu^{\prime \prime}=0$, $N^{(3)}=0$,
$N^{(6)}=M^{(6)}$, $p_m=0$. Note that in this limit (\ref{it}) is
trivially satisfied.
An example of a solution yielding stabilized dimensions and three dimensions
growing large corresponds to $k_1=\frac{1}{9}$ and $k_2=\frac{9}{7}$.
The numerical solution of \cite{stable} and the analytical approximation
used in this paper are compared in Fig. 1, for the above values of the
constants $k_1$ and $k_2$.

\begin{figure}[!]
\includegraphics[totalheight=6 in,keepaspectratio]{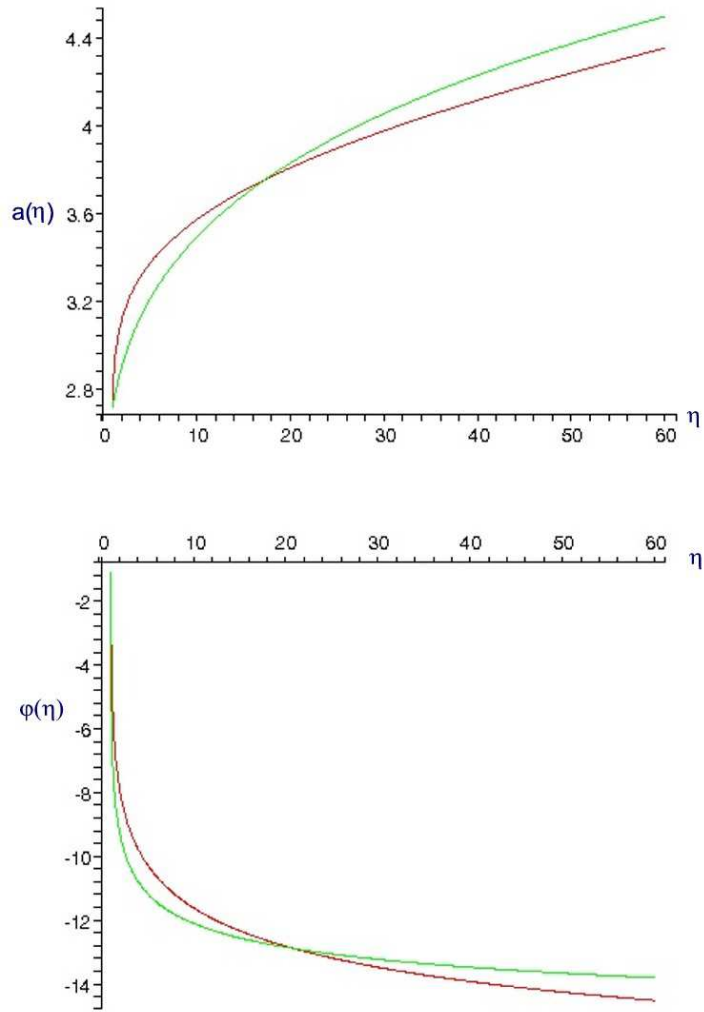}
\caption{A comparison between the numerical background solutions
obtained in \cite{stable} (red or light line) and the analytical
approximation used in this paper (green or dark line).}
\label{fig1}
\end{figure}

\section{Scalar Metric Perturbations}

In this section we consider the growth of scalar metric perturbations
(see e.g. \cite{MFB} for a comprehensive review of the theory of
cosmological perturbations) due to the presence of string inhomogeneities.
We are interested in the case where the fluctuations depend only
on the external coordinates and conformal time, not on the coordinates
of the internal dimensions. For simplicity we work in the generalized
longitudinal gauge in which the metric perturbations are only in
the diagonal metric elements \footnote{As discussed e.g. in
\cite{Dorca}, for scalar perturbations depending on all spatial coordinates
it would be inconsistent to choose the perturbed metric completely
diagonal, and one would have to add a metric coefficient to the
$dt dx^m$ terms, where $x^m$ are the coordinates of the internal
dimensions. However, as discussed in \cite{GG96}, if the fluctuations
are independent of the coordinates $x^m$, as in our case, the coefficient
can be chosen to vanish, and thus the perturbed metric is completely
diagonal.}. Thus, the metric including linear fluctuations is given by
\be \label{pertmetric}
ds^{2}=e^{2 \lambda(\eta)} \Bigl( (1+2\phi) d\eta^{2}- (1-2\psi)\delta_{i j}
dx^{i} dx^{j}\Bigr) -e^{2 \nu(\eta)}(1-2\xi) \delta_{m n}
dx^{m}dx^{n}.
\ee
The dilaton $\varphi$ also fluctuates about its background
value $\varphi_0$. The dilaton fluctuation $\chi$ is determined by
\be
\varphi = \varphi_0 + \delta \varphi \;\;\;\;\;\; \chi \equiv
\delta \varphi.
\ee
In the above, the fluctuating fields $\chi, \phi, \psi$ and $\xi$ are
functions of the external coordinates $x^i$ and time, i.e.
\be
\chi=\chi(\eta,x^i), \: \: \: \: \phi=\phi(\eta,x^i), \: \: \: \:  \psi=\psi(\eta,x^i),\: \: \: \:
\xi=\xi(\eta,x^i).
\ee

The perturbations of the matter energy momentum tensor result
from over-densities and under-densities in the number of strings.
From (\ref{sources1})-(\ref{sources1b}) and noting
that we are interested in the case when $N^{(3)}=0$ and
$M^{(6)}=N^{(6)}$ we find,
\bea &\delta \rho=\delta \rho_{{\text w}} + \delta \rho_{{\text
m}},\\ &\delta \rho_{{\text w}}=30 \mu N \xi e^{-3\lambda}+18 \mu
N \psi e^{-3\lambda}+6 \delta N^{(6)}e^{-3\lambda},
\\ &\delta \rho_{{\text m}}=42 \mu N \xi e^{-3\lambda}+18 \mu M \xi e^{-4\lambda}+
18 \mu N \psi e^{-3\lambda}+12 \mu M \psi e^{-4\lambda}+6 \mu
\delta M^{(6)}e^{-3\lambda}+\\&+3 \mu \delta M e^{-4\lambda} ,
\\ &\delta p_{\lambda}=6 \mu M \xi e^{-4 \lambda}+4 \mu M \psi e^{-4 \lambda}+
\mu \delta M e^{-4 \lambda},\\
&\delta p_{\nu}=2 \mu N \xi e^{-3 \lambda}-\mu \delta N^{(6)}
e^{-3 \lambda}+\mu \delta M^{(6)} e^{-3 \lambda}, \eea
where we define\footnote{Notice that we must be careful to
distinguish between the perturbed quantities $\delta N^{(6)}$ and
$\delta M^{(6)}$.} $N \equiv N^{(6)}=M^{(6)}$ and $M \equiv
M^{(3)}$. The fluctuations $\delta N^{(6)}$, $\delta M^{(6)}$, and
$\delta M$ are taken as functions of both conformal time and the
external space, e.g. $\delta N^{(6)}=\delta N(\eta, x^i)$.

It follows from (\ref{theconservationeq}) that the perturbed sources
obey modified conservation equations for both the winding and momentum modes,
\be \label{pce}
\delta \rho^{\prime \: {\text w,m}} +\sum_{i=1}^9 \lambda_{i}^{\prime} (\delta \rho^{\text w,m}
- \delta p^{\text w,m}_i)+\sum_{i=1}^9 \delta \lambda_i^{\prime} (\rho^{\text w,m}
- p^{\text w,m}_i)=0,
\ee
where $\delta \lambda=a^{-1} \delta a =-\psi$ and $\delta
\nu=b^{-1} \delta b=-\xi$ are spatial variations.

We rewrite (\ref{eomb}) to take the more familiar form of the
Einstein and dilaton equations, namely

\bea
R_{\mu}^{\nu}-\frac{1}{2} \delta_{\mu}^{\nu} R=e^{2\varphi} T_{\mu}^{\nu}-2 g^{\alpha \nu}
\nabla_{\mu} \nabla_{\alpha} \varphi + 2 \delta_{\mu}^{\nu} \Bigl( g^{\mu \nu} \nabla_{\mu} \nabla_{\nu} \varphi
-g^{\mu \nu} \partial_{\mu} \varphi \partial_{\nu} \varphi
\Bigr),  \nonumber \\
g^{\mu \nu} \partial_{\mu} \varphi \partial_{\nu} \varphi
-\frac{1}{2} g^{\mu \nu} \nabla_{\mu} \nabla_{\nu}\varphi=\frac{1}{4}
e^{2 \varphi} T^{\mu}_{\mu},
\eea
where we invoke Planckian units (i.e. $\gn= 1$).
Plugging the perturbed metric (\ref{pertmetric}) and dilaton into
these equations, making use of the background equations of motion,
and linearizing the equations about the background (i.e. keeping
only terms linear in the fluctuations) yields the following set of
equations:
\bea \label{eq111}
{\vec \nabla}^2 \psi +3 {\vec \nabla}^2\xi -9 \ch \xp -3\ch \pp -3
\ch^2 \phi =\frac{1}{2} e^{2\varphi + 2\lambda}\Bigl( 2\chi T^0_0 +\delta T^0_0 \Bigr)
-6 \ch \phi \vp \nonumber \\-3 \pp \vp-6\xp \vp -{\vec \nabla}^2 \chi +3 \ch
\cp +2 \phi \varphi^{\prime 2}-2\cp \vp,
\eea
\be
\partial_{i}\pp+3\partial_{i}\xp+\ch\partial_{i}\phi-3\ch\partial_{i}\xi=
\frac{1}{2} e^{2\varphi+2\lambda} \delta T_{0}^{i}+\partial_{i}\phi
\vp-\partial_{i}\cp+\ch\partial_{i}\chi,
\ee
\be \label{usefull}
\partial_{i}\partial_{j} \Bigl( \phi-\psi-6\xi-2\chi \Bigr)=0 \;\;\;\;\;\; i \neq j,
\ee
\begin{widetext}
\bea
\Bigl( \partial_{i}^{2}-\vec{\nabla}^{2} \Bigr)\Bigl( \phi-\psi-6\xi
\Bigr)-2\ppp-6\xpp-
4\ch\pp -6\ch\xp-2\ch{^2}\phi-4\ch^{\prime}\phi-2\fp\ch
\nonumber \\ =e^{2 \varphi+2\lambda} \Bigl( 2\chi T^{i}_{i}+\delta T^{i}_{i} \Bigr)
+2\partial_{i}^{2}\chi-4\phi\vpp-2\fp\vp  -4\ch\phi\vp-4\pp\fp-12\xp\vp\nonumber\\
+2\cpp-2\vec{\nabla}^{2}\chi+2\ch\cp+4\phi\varphi^{\prime
2}-4\cp\vp,
\eea
\end{widetext}
\bea
-{\vec \nabla}^2 \phi +5 {\vec \nabla}^2 \xi -5 \xpp +2 {\vec \nabla}^2
\psi-3 \ppp-10 \ch \xp -3 \fp \ch -9 \ch \pp -6 \ch^2 \phi -6
\ch^{\prime} \phi \nonumber \\ =
 e^{2\varphi+2\lambda} \Bigl( 2 \chi T^m_m + \delta T^m_m \Bigr)
 -4 \phi \vpp -2 \fp \vp -8 \ch \phi \vp -6 \pp \vp -10 \xp \vp  +2
 \cpp \nonumber \\-2{\vec \nabla}^2 \chi +4 \ch \cp +4 \phi \varphi^{\prime 2}
 -4 \cp \fp,
\eea
\bea \label{eq113}
-2 \phi \varphi^{\prime 2}+2 \vp \cp+\phi \vpp -6 \psi \ch
\vp+\frac{1}{2} \fp \vp -2 \ch \phi \vp - \frac{3}{2} \pp \vp -3
\xp \vp -\frac{1}{2}\cpp \nonumber \\ +\frac{1}{2}{\vec \nabla}^2
\chi -\ch \cp
=\frac{1}{4}e^{2\varphi + 2\lambda} \Bigl( 2\chi T + \delta T
\Bigr),
\eea
where $T\equiv T^{\mu}_{\mu}$ is the trace.
The modified conservation equations (\ref{pce}) take the form
\bea \label{yep1b}
&\frac{d }{d \eta}\Bigl( \delta N^{(6)} \Bigr)=7 N \xp,\\
&42 \mu N \xp-72 \mu M \xi \lp e^{- \lambda}-48 \mu M \psi \lp e^{- \lambda}
+12 \mu M \pp e^{- \lambda}+6 \mu \frac{d }{d \eta}\Bigl( \delta M^{(6)}\Bigr)\nonumber \\&
-12 \mu \delta M \lp e^{- \lambda} +3 \mu \frac{d }{d \eta}\Bigl( \delta M \Bigr) e^{-
\lambda}=0.
\label{yep1} \eea
These equations give us the evolution of the metric perturbations
$\phi$, $\psi$, and $\xi$ in terms of the matter perturbations
$\chi$, $\delta \rho$, and $\delta p_i$.  At first glance, it may
appear that the above system is over-determined since we have
eight equations for seven unknowns.  However, as is the case in
standard cosmology, the conservation equations are not independent
of the Einstein equations. Thus, we can choose to keep only one
of the modified conservation equations and our system will be
consistent.

\section{Evolution of Fluctuations}

In this section we focus on the evolution of long wavelength perturbations,
since instabilities on these scales would be the most dangerous for the
success of BGC at providing a mechanism which allows exactly three
spatial dimensions to become large and stabilizes the radius of the other
dimensions at a microscopic value. We focus on the time interval $\eta \gg 1$
when the hierarchy in scales between the large and the small spatial
dimensions has already developed, and where we can use the
analytical approximations to the background dynamics given in Section II.

Since we are focusing on long wavelength fluctuations, we can neglect
all terms with spatial gradients in Equations (\ref{eq111})-(\ref{eq113}).
Making use of (\ref{usefull}) we
can eliminate one of the scalar metric perturbations, $\phi$, from the rest of
the Equations (\ref{eq111})-(\ref{eq113}).  We take Equation (\ref{eq111})
as a constraint on the initial data, noticing that it only contains first
derivatives of the perturbation variables.
By using the background solution (\ref{backsoln}) we find that the remaining
equations represent a coupled system of harmonic oscillators with
time dependent coefficients. With these approximations, the equations become
\footnote{The full equations can be found in the Appendix.}
\bea
&-2\cpp-2\ppp-6\xpp-\frac{230}{7 \eta} \xp -\frac{176}{21
\eta}\pp-\frac{230}{21 \eta} \cp -\frac{12868}{3969 \eta^2} \chi
-\frac{6434}{3969 \eta^2} \psi-\frac{12868}{1323 \eta^2}\xi=0,\\
&-3\ppp -5\xpp -2\cpp -\frac{244}{21 \eta} \pp -\frac{1978}{63 \eta}\xp -\frac{718}{63 \eta}\cp
-\frac{2672}{1323 \eta^2}\psi-\frac{5344}{441 \eta^2}
\xi -\frac{5344}{1323 \eta^2}\chi=0, \\
&-\frac{1}{2}\cpp-\frac{250}{63 \eta}\cp+\frac{9}{7 \eta} \pp
-\frac{170}{49 \eta^2}\chi-\frac{43}{49 \eta^2}\psi
-\frac{510}{49 \eta^2}\xi=0.
\eea

Notice that the string source terms, $\delta N$ and $\delta M$,
have disappeared from the equations. This is due to the fact
that, for long wavelength fluctuations, the terms due to purely
gravitational dynamics (e.g. the damping terms in the evolution of the
radius of the internal dimensions as a consequence of the expansion
of the large dimensions) are more important than the matter
sources. This is not too surprising based on the results of the
theory of linear cosmological perturbations \cite{MFB} which show
that on super-Hubble scale matter is pulled along by gravity, but that
it is the purely gravitational dynamics which determines the
growth rate of the fluctuations. In analogy, we find here that the
terms representing the string matter sources are
sub-leading in the equations (see Appendix). This realization is
the key physical reason which leads to our ultimate conclusion that
also in BGC there are no instabilities of long wavelength fluctuations.

The above set of coupled equations can be solved exactly, and we find
the leading terms to be
\bea
\psi \sim \frac{c_1}{\eta^{0.06}}+\frac{c_2}{\eta^{0.36}},\\
\xi \sim c_3+\frac{c_4}{\eta^{0.06}},\\
\chi \sim c_5+\frac{c_6}{\eta^{0.06}}.
\eea
This is our main result.  First, we have found that the
perturbation of the internal geometry $\xi$ is constant to leading
order.  Thus, the stabilization mechanism resulting from
considering winding and momentum modes remains viable in the
presence of string inhomogeneities and dilaton fluctuations.
Secondly, the geometry of the $3+1$ dimensions growing large does
not suffer from exponential instabilities.  In fact, we have power-law decay
for $\psi$ and the dilaton perturbation $\chi$ decays to a constant.

The crucial point that led to this behavior was the
fact that the string source terms were sub-leading in the
equations above.  Physically this means that the number of winding and momentum modes
moving in or out of a region is negligible compared to the expansion rate and the
evolution of the perturbations.  Therefore, there is an {\em
averaging} of the long wavelength perturbations that leads to a
smoothing effect and the string inhomogeneities are rendered
harmless.

\section{Conclusions}

We have considered the evolution of linearized string inhomogeneities and
dilaton fluctuations in the background of Brane Gas Cosmology (BGC).
We considered fluctuations which are independent of the spatial
coordinates of the internal dimensions. We
have derived the perturbation equations for BGC in the longitudinal gauge
and solved them in the long wavelength approximation.
We have found that for long wavelengths, the effects of the string
sources is subleading compared to the purely gravitational and dilaton
terms in the fluctuation equations. This generalizes the result of
the conventional theory of cosmological perturbations in Einstein
gravity in four space-time dimensions which states that on scales
larger than the Hubble radius, self-gravitational effects dominate
the dynamics, and that the matter sources are simply dragged along
by the metric. In particular, there are no instabilities of the
background solution of BGC towards such long wavelength fluctuations.
Thus, we find that the predictions of BGC are robust
towards the effects of matter and dilaton fluctuations.  In particular,
the stabilization mechanism for the extra
dimensions (volume modulus) remains operative in the presence of
these inhomogeneities, while the $3+1$ dimensional space-time continues to grow
without instabilities.

Although this is a promising result for BGC, there are still many
questions to be answered.  Our analysis has been limited to the
linear regime, while non-linear effects may be crucial. Our
linearized analysis corresponds to taking the localized string
sources and smearing them out over a length scale larger than the
Hubble radius to obtain a string gas description. It would
be interesting to see if these results hold when considering
individual strings and their interactions and in what limits the
string gas approximation holds.  Also, BGC offers a dynamical way
to stabilize the volume modulus, but we have not addressed the
issue of stabilizing the dilaton and we have also ignored the role
of fluxes.  We leave these questions for further investigations.

\begin{acknowledgments}

SW would like to thank Alan Guth and Steve Gubser for useful discussions.
RB was supported in part by the U.S. Department of Energy under
Contract DE-FG02-91ER40688, TASK A. SW was supported in part by
the NASA Graduate Student Research Program.

\end{acknowledgments}

\appendix
\section{Perturbation equations}
In this Appendix we present a detailed derivation of the solutions
presented in section 4. In the long wavelength limit we can neglect gradient terms and the equations
(\ref{eq111})-(\ref{eq113}) reduce to
\bea
\label{eqn00}
&-9\ch\xp -3 \ch \pp -3\ch^2 \phi
=\frac{1}{2} e^{2\lambda+2 \varphi} \Bigl( 2\chi T^0_0 + \delta T^0_0  \Bigr)
-6\ch\phi \vp -3 \pp \vp-6 \xp \vp+\nonumber \\
&+3\ch\cp +2 \phi \varphi^{\prime 2} - 2 \cp \vp,
\eea
\be
\label{eqnij}
\partial_{i}\partial_{j}\Bigl( \phi-\psi-6\xi-2\chi \Bigr)=0
\;\;\;\;\;\; i \neq j,
\ee
\bea
\label{eqnii}
&-2 \ppp -6 \xpp -4\ch\pp -6\ch\xp -2\ch^2\phi
-4\ch^{\prime}\phi -2\fp \ch
=e^{2\lambda+2 \varphi} \Bigl(
 2\chi T^i_i +\delta T^i_i \Bigr)+ \nonumber \\
 &-4 \phi \vpp -2 \fp \vp -4\ch\phi
 \vp -4 \pp \vp-12 \xp \vp + 2 \cpp +2\ch\cp
  +4 \phi \varphi^{\prime 2} -4 \cp \vp,
\eea
\bea
\label{eqnmm}
&-5\xpp -3\ppp-10\ch\xp-3\fp \ch -9\ch\pp -6\ch^2\phi
-6H^{\prime} \phi =e^{2\lambda+2 \varphi} \Bigl(
2\chi T^m_m +\delta T^m_m \Bigr), \nonumber \\
&-4 \phi \vpp -2\fp \vp -8\ch \phi
\vp -6\pp\vp-10\xp\vp+2\cpp+4\ch\cp+4\phi\varphi^{\prime
2}-4\cp\vp,
\eea
\bea
\label{eqndilaton}
&-2\phi\varphi^{\prime
2}+2\vp\cp+\phi\vpp-6\psi\ch\vp+\frac{1}{2}\vp\fp-2\ch\phi\vp-\frac{3}{2}\pp\vp-3\xp\vp-\frac{1}{2}\cpp-\ch\cp
\nonumber \\
&=\frac{1}{4} e^{2\lambda+2\varphi}\Bigl( 2\chi\mu T^{\mu}_{\mu}+\delta T^{\mu}_{\mu}
\Bigr).
\eea

Making use of the background solution (\ref{backsoln}) and
(\ref{eqnij}) to eliminate the scalar metric perturbation $\phi$, we find
the following set of equations describing the evolution;
\bea \label{gig1}
&-\frac{61}{21 \eta}\cp -\frac{61}{7 \eta}\xp-\frac{88}{21 \eta} \pp
+ \Biggl( -\frac{11114}{1323 \eta^2}-\frac{3000}{\eta^{\frac{176}{63}}}-\frac{12000}{\eta^{\frac{169}{63}}}
\Biggr)\chi
+ \Biggl( -\frac{36000}{\eta^{\frac{169}{63}}}-\frac{9000}{\eta^{\frac{176}{63}}}-\frac{11114}{441 \eta^{2}}
\Biggr)\xi+ \nonumber \\
&+ \Biggl( -\frac{6000}{\eta^{\frac{176}{63}}}-\frac{5557}{1323 \eta^{2}}-\frac{18000}{\eta^{\frac{169}{63}}}
\Biggr)\psi-\frac{3}{\eta^{\frac{169}{63}}}\delta
M^{(6)}-\frac{3}{\eta^{\frac{169}{63}}}\delta
N^{(6)}-\frac{3}{2 \eta^{\frac{176}{63}}}\delta M^{(3)}=0, \\
&-2\cpp-2\ppp-6\xpp-\frac{230}{7 \eta} \xp -\frac{176}{21
\eta}\pp-\frac{230}{21 \eta} \cp +\Biggl( -\frac{2000}{\eta^{\frac{176}{63}}}-\frac{12868}{3969 \eta^2} \Biggr)\chi
+\Biggl( -\frac{4000}{\eta^{\frac{176}{63}}}-\frac{6434}{3969 \eta^2} \Biggr)\psi+
\nonumber \\
&+\Biggl( -\frac{6000}{\eta^{\frac{176}{63}}}-\frac{12868}{1323 \eta^2}
\Biggr)\xi-\frac{1}{\eta^{\frac{176}{63}}}\delta M^{(3)}=0,
\\
&-3\ppp -5\xpp -2\cpp -\frac{244}{21 \eta} \pp -\frac{1978}{63 \eta}\xp -\frac{718}{63 \eta}\cp
-\frac{2672}{1323 \eta^2}\psi+\Biggl( -\frac{2000}{\eta^{\frac{169}{63}}}-\frac{5344}{441 \eta^2}\Biggr)
\xi +\nonumber \\ &-\frac{5344}{1323 \eta^2}\chi+\frac{1}{\eta^{\frac{169}{63}}}\delta N^{(6)}
-\frac{1}{\eta^{\frac{169}{63}}}\delta M^{(6)}=0, \\
&-\frac{1}{2}\cpp-\frac{250}{63 \eta}\cp+\frac{9}{7 \eta} \pp
+\Biggl( -\frac{170}{49 \eta^2}-\frac{6000}{\eta^{\frac{169}{63}}}
\Biggr)\chi +\Biggl( -\frac{9000}{\eta^{\frac{169}{63}}}-\frac{43}{49 \eta^2} \Biggr)\psi
+\Biggl( -\frac{510}{49 \eta^2}-\frac{15000}{\eta^{\frac{169}{63}}} \Biggr)\xi+
\nonumber \\&-\frac{3}{\eta^{\frac{169}{63}}}\delta N^{(6)} =0.
\label{gig2}
\eea

The first equation, containing only first derivatives, is taken as
a constraint on the initial data.  We are interested in the late time
behavior, i.e. $\eta \gg 1$.  Keeping only the leading order terms
the remaining equations can be approximated as,
\bea
&-2\cpp-2\ppp-6\xpp-\frac{230}{7 \eta} \xp -\frac{176}{21
\eta}\pp-\frac{230}{21 \eta} \cp -\frac{12868}{3969 \eta^2} \chi
-\frac{6434}{3969 \eta^2} \psi-\frac{12868}{1323 \eta^2}\xi=0,\\
&-3\ppp -5\xpp -2\cpp -\frac{244}{21 \eta} \pp -\frac{1978}{63 \eta}\xp -\frac{718}{63 \eta}\cp
-\frac{2672}{1323 \eta^2}\psi-\frac{5344}{441 \eta^2}
\xi -\frac{5344}{1323 \eta^2}\chi=0, \\
&-\frac{1}{2}\cpp-\frac{250}{63 \eta}\cp+\frac{9}{7 \eta} \pp
-\frac{170}{49 \eta^2}\chi-\frac{43}{49 \eta^2}\psi
-\frac{510}{49 \eta^2}\xi=0.
\eea

This set of coupled equations can be solved exactly and we find
the leading terms to be
\bea
\psi \sim \frac{c_1}{\eta^{0.06}}+\frac{c_2}{\eta^{0.36}},\\
\xi \sim c_3+\frac{c_4}{\eta^{0.06}},\\
\chi \sim c_5+\frac{c_6}{\eta^{0.06}}.
\eea

As a consistency check, we can plug this result back into the original equations
(\ref{gig1})-(\ref{gig2}).  In addition we must consider the
evolution of $\delta N^{(6)}$, $\delta M^{(6)}$ and $\delta M$ given
by (\ref{yep1b}) and (\ref{yep1}).
We find that it was consistent to neglect
the time evolution of the string sources (i.e. $\delta N$ and
$\delta M$) compared with the expansion of the background and the evolution
of the perturbations.  In addition, we find that at late times this holds as an exact
solution of the perturbation equations.  In this way, we find a
self consistent solution for the behavior of the long wavelength
linear perturbations.


\end{document}